\documentclass[twocolumn,aps,superscriptaddress]{revtex4}
\usepackage{graphicx}
\usepackage{float}
\usepackage{dcolumn}
\usepackage{bm}
\usepackage{color}
\usepackage{amsmath}
\usepackage{amssymb}
\usepackage{amsfonts}
\usepackage{esint}
\usepackage{times}
\usepackage{xcolor}
\usepackage{phaistos}
\usepackage{braket}
\usepackage{comment}

\usepackage{pdfpages}

\usepackage[colorlinks,linkcolor=blue,anchorcolor=blue,citecolor=blue]{hyperref}

\begin{document}

\title{Spurious microwave crosstalk in floating superconducting circuits}

\author{Peng Zhao}
\email{shangniguo@sina.com}
\affiliation{Beijing Academy of Quantum Information Sciences, Beijing 100193, China}
\author{Yingshan Zhang}
\affiliation{Beijing Academy of Quantum Information Sciences, Beijing 100193, China}
\author{Xuegang Li}
\affiliation{Beijing Academy of Quantum Information Sciences, Beijing 100193, China}
\author{Jiaxiu Han}
\affiliation{Beijing Academy of Quantum Information Sciences, Beijing 100193, China}
\author{Huikai Xu}
\affiliation{Beijing Academy of Quantum Information Sciences, Beijing 100193, China}
\author{Guangming Xue}
\email{xuegm@baqis.ac.cn}
\affiliation{Beijing Academy of Quantum Information Sciences, Beijing 100193, China}
\author{Yirong Jin}
\affiliation{Beijing Academy of Quantum Information Sciences, Beijing 100193, China}
\author{Haifeng Yu}
\affiliation{Beijing Academy of Quantum Information Sciences, Beijing 100193, China}

\date{\today}

\begin{abstract}
Crosstalk is a major concern in the implementation of large-scale
quantum computation since it can degrade the performance of qubit addressing
and cause gate errors. Finding the origin of crosstalk and separating
contributions from different channels are essential prerequisites for figuring out
crosstalk mitigation schemes. Here, by performing
circuit analysis of two coupled floating transmon qubits, we demonstrate that, even if
the stray coupling, e.g., between a qubit and the drive line of its
nearby qubit, is absent, microwave crosstalk between qubits can still
exist due to the presence of a spurious crosstalk channel. This channel arises
from free modes, which are supported by the floating structure of transmon qubits, i.e., the
two superconducting islands of each qubit with no galvanic connection
to the ground. For various geometric layouts of floating transmon qubits,
we give the contributions of microwave crosstalk from the spurious channel and
show that this channel can become a performance-limiting factor in qubit addressing.
This research could provide guidance for suppressing microwave crosstalk between
floating superconducting qubits through the design of qubit circuits.

\end{abstract}

\maketitle


\section{Introduction}

Integrating a growing number of qubits without scarifying quantum gate
performance is a key task in the implementation of large-scale quantum
computers with superconducting qubits \cite{Martinis2015}. One of the main
obstacles that needs to be overcome, in this task, is crosstalk, including classical
crosstalk due to unintended classical electromagnetic
couplings \cite{Wenner2011,Martinis2014,Rosenberg2019,Patterson2019,Huang2021,Abrams2019,Dai2021} and
quantum crosstalk arising from residual quantum coupling \cite{Patterson2019,Gambetta2012,Mundada2019,Wei2021,Zhao2022}, which
can make qubit addressing a challenge \cite{Gambetta2012} and degrade gate performance in
multiqubit quantum processors \cite{Sarovar2020}. In this context, the progress in understanding
and mitigating crosstalk has made indispensable contributions to the impressive
achievements toward developing large-scale superconducting quantum
computing over the past decade.

\begin{figure}[tbp]
\begin{center}
\includegraphics[keepaspectratio=true,width=\columnwidth]{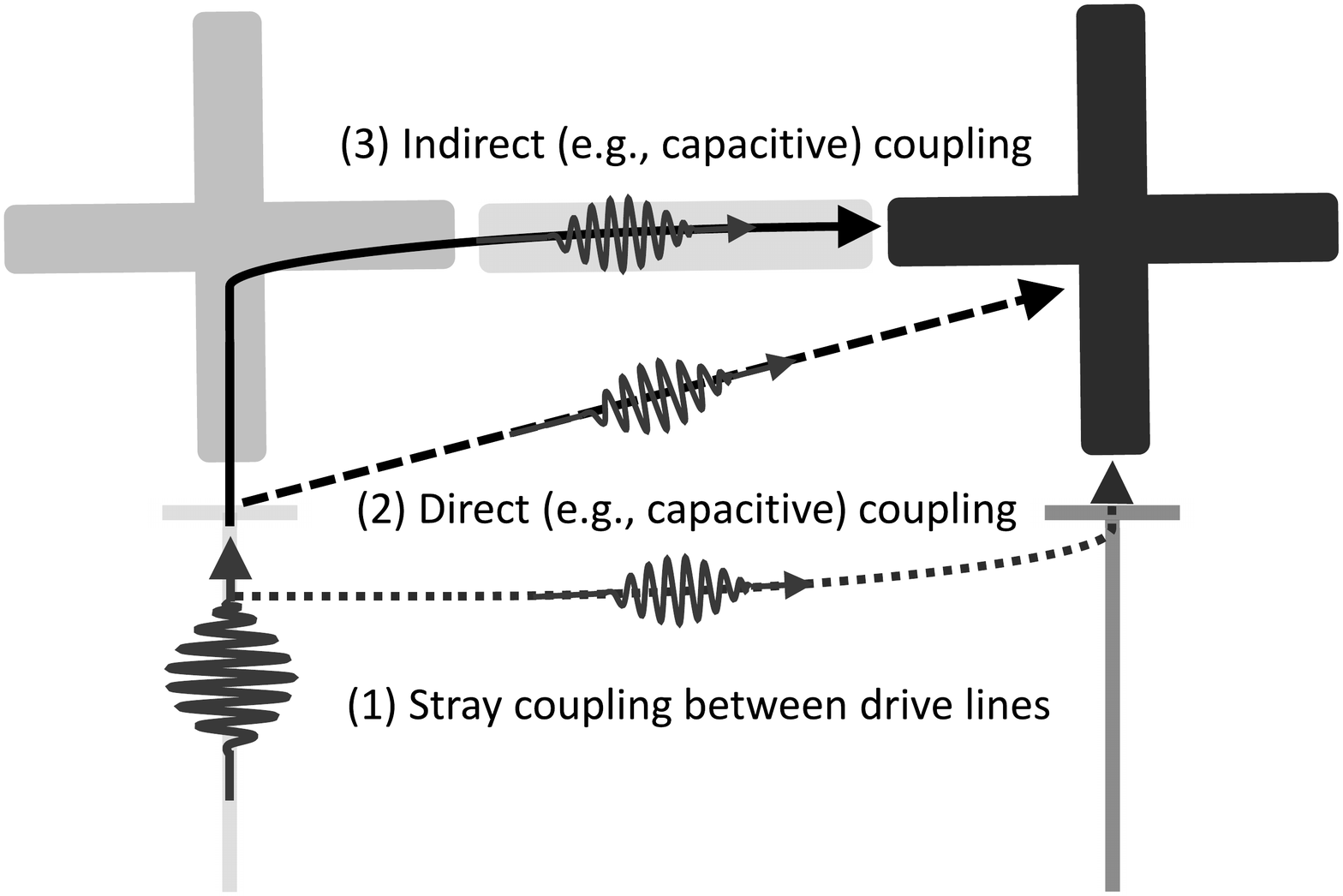}
\end{center}
\caption{Three possible channels of microwave crosstalk between two
coupled superconducting qubits. Each qubit (cross-shaped) has a dedicated drive
line (T-shaped) for single-qubit addressing. The two qubits are coupled via a
coupler circuit (strip-shaped), such as a direct coupling capacitor or a bus coupler. Microwave
crosstalk can be arisen from: (Type-1) the stray coupling between
the two drive lines; (Type-2) the direct or (Type-3) indirect coupling between one qubit and the drive line
of its neighbor.}
\label{fig1}
\end{figure}

One of the most ubiquitous crosstalk for superconducting quantum
processors is the microwave crosstalk, which describes that
microwave drives applied to one qubit can cause unintended drives
felt by the others. To address this issue, various active cancelation
methods, i.e., one first characterizes it and then cancels it actively with
a compensation drive, have been demonstrated \cite{Sung2021,Nuerbolati2022}.
Nevertheless, mitigating the crosstalk at device level may
complement existing active approaches and further reduce the needed physical
resource, especially for large-scale quantum processors. Indeed, previous
works show that the crosstalk can be suppressed at the device level, but
only if the origin of the crosstalk and contributions from different crosstalk
channels are well understood \cite{Wenner2011,Huang2021}.

Generally, to implement universal control over qubits in
quantum processors, at least a single drive
line (drive channel) per qubit is needed. Hence, as shown in Fig.~\ref{fig1},
there are three possible microwave crosstalk channels in superconducting quantum
processors, including (Type-1) stray coupling between the dedicated drive lines
of qubits, the direct (Type-2) and the indirect (Type-3) coupling between one qubit
and the drive line of others. For large-scale quantum processors, a high
density of control wiring is required, thus the crosstalk through
the first two channels could become more serious \cite{Wenner2011,Rosenberg2019,Huang2021}. Generally,
the two channels can be mitigated by improving the physical
isolation between drive lines and qubits \cite{Rosenberg2019}. For the Type-3 channel, however, its
physical origin and mitigation seems to be more nontrivial. In principle, its origin
can be modeled by the indirect coupling between one qubit and the
drive line of the others through an effective circuit network \cite{Solgun2019}. Physically, the
circuit network could arise from the presence of package modes
or chip modes \cite{Wenner2011,Rosenberg2019}, and could even relate to the
qubit itself \cite{Johnson2011,Galiautdinov2012}. Nevertheless, the exact nature of this
circuit network and its contribution to the microwave crosstalk are less studied.

In this work, we present a spurious microwave crosstalk channel (Type-3) enabled by the presence of
free modes in floating transmon circuits \cite{Kerman2020,Ding2021,Long2020,Koch2007}. By performing
circuit analysis of coupled floating transmon qubits with different geometric
layouts \cite{Paik2020,Rahamim2017,Mamin2022}, we show that the microwave crosstalk
contributed from this spurious channel can become non-negligible, thus potentially
limiting the performance of qubit addressing. More importantly, this
crosstalk channel only depends on the qubit circuit
itself, thus acting as an intrinsic channel, which can exist even when
the stray coupling, e.g., between drive lines and qubits, is absent. This feature also suggests
that this spurious channel can be mitigated through qubit circuit design.

This paper is organized as follows. In Sec.~\ref{SecII}, we analyze the quantum
circuit of two direct-coupled floating transmon qubits (in the Appendix we
further extended our analysis to the case, where floating transmon qubits are
coupled via a grounded or floating bus coupler) and show that the presence of the
free modes can induce a spurious microwave crosstalk channel. In Sec.~\ref{SecIII}, for
transmon qubits with different qubit geometric layouts, we give the
contributions of microwave crosstalk from the spurious channel. In Sec.~\ref{SecIV}, we
give discussions on the relation of the free-mode-mediated spurious crosstalk
channel in our work with the free-mode mediated inter-qubit interactions illustrated in two
recent works \cite{Sete2021,Yanay2022} and show that the present work can be viewed as
complements to the two earlier works, extending the free-mode mediated interactions
from "quantum regime" (for inter-qubit coupling) to "semi-classical regime" (for
classical microwave crosstalk). Finally, in Sec.~\ref{SecV}, we provide
a summary of our work.

\section{spurious microwave crosstalk channel}\label{SecII}

\begin{figure}[tbp]
\begin{center}
\includegraphics[keepaspectratio=true,width=\columnwidth]{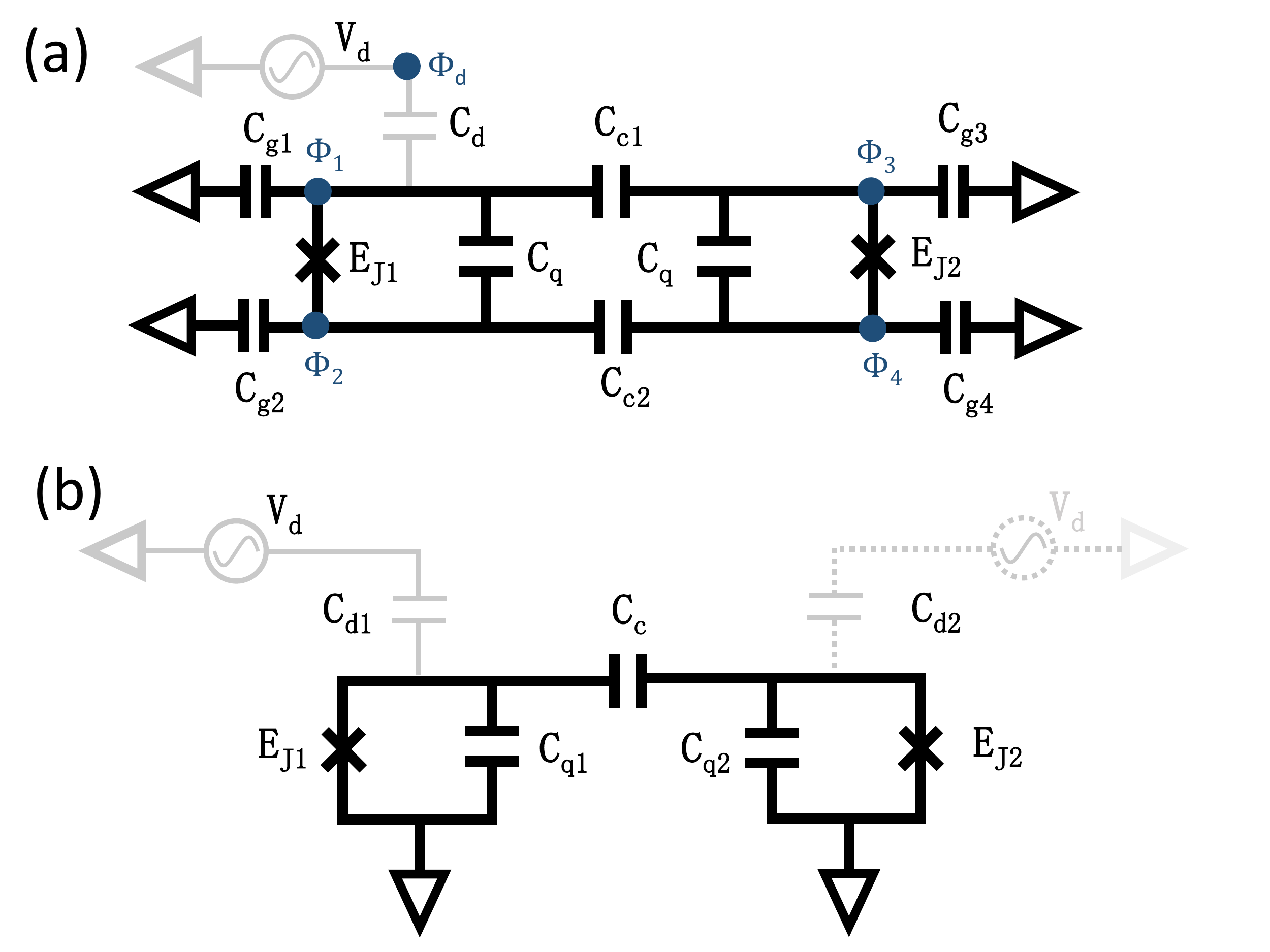}
\end{center}
\caption{(a) Schematic circuit diagram for two direct-coupled floating transmon qubits ($Q_{1}$ and $Q_{2}$), where a dedicated
drive line (left) is coupled capacitively to one of the two qubits, i.e., $Q_{1}$. After removing the free
modes supported by the floating structure of qubits, the circuit can be transformed into an equivalent circuit shown in (b), where two
grounded tranomon qubits are coupled capacitively. Here, the virtual drive line (right, grey dashed line) is introduced
to model the crosstalk due to the presence of a spurious channel.}
\label{fig2}
\end{figure}

To understand the origin of the spurious microwave crosstalk channel, we consider
a superconducting circuit comprising two direct-coupled floating transmon qubits ($Q_{1}$ and $Q_{2}$), and one
of the two qubits ($Q_{1}$) is coupled capacitively to a voltage source, as shown
in Fig.~\ref{fig2}(a). When expressed in terms of the node flux variables $\Phi_{j}$ and $\phi_{j}=2\pi\Phi_{j}/\Phi_{0}$
with $\Phi_{0}$ the magnetic flux quantum, the circuit Lagrangian is given by (details on its
derivation can be found in Appendix~\ref{A}) \cite{Yanay2020,Vool2017}
\begin{equation}
\begin{aligned}\label{eq1}
&\mathcal{L}=\frac{1}{2}\mathbf{\dot{\Phi}}^{T} \mathbf{C}\mathbf{\dot{\Phi}}+E_{J1}\cos(\phi_{1m})+E_{J2}\cos(\phi_{2m}),
\end{aligned}
\end{equation}
where $\mathbf{\Phi}=(\Phi_{d}\,\Phi_{1p}\,\Phi_{1m}\,\Phi_{2p}\,\Phi_{2m})^{T}$ with $\Phi_{1p(m)}=\Phi_{1}\pm\Phi_{2}$ and $\Phi_{2p(m)}=\Phi_{3}\pm\Phi_{4}$, $E_{J1}$ and $E_{J2}$ are the Josephson energies, and $\mathbf{C}$ denotes the
capacitance matrix of the circuit. Accordingly, the circuit Hamiltonian can be constructed
as $H=\sum_{j}Q_{j}\dot{\Phi}_{j}-E_{J1}\cos(\phi_{1m})-E_{J2}\cos(\phi_{2m})$
with the charge variables $Q_{j}=\partial\mathcal{L}/\partial \dot{\Phi}_{j}$.
Here, the modes associated with the variables $Q_{1m}$ and $Q_{2m}$ are the qubit modes, which have both the charge energy
and the potential energy \cite{Koch2007}, while the modes associated with $Q_{1p}$ and $Q_{2p}$, and which don't have
potential terms in the Hamiltonian, are the free modes \cite{Kerman2020,Ding2021,Long2020}.
From the derivation and also mentioned in previous works \cite{Kerman2020,Ding2021,Long2020}, the presence of the
free modes are supported by the floating structure of transmon qubits, i.e., the two superconducting islands
of the qubit have no galvanic connection to the ground, as shown in Fig.~\ref{fig2}(a), contributing
to an additional quantum degree of freedom.

Since free modes actually don't participate in the circuit
dynamics \cite{Kerman2020,Ding2021,Long2020}, one can drop the charge terms
corresponding to the two free modes and rewrite the circuit Hamiltonian as
\begin{equation}
\begin{aligned}\label{eq2}
&H_{r}=\frac{1}{2}\mathbf{Q_{r}}^{T} \mathbf{C_{r}}^{-1}\mathbf{Q_{r}}-E_{J1}\cos(\phi_{1m})-E_{J2}\cos(\phi_{2m}),
\end{aligned}
\end{equation}
with $\mathbf{Q_{r}}=(Q_{d}\,Q_{1m}\,Q_{2m})^{T}$.
Here, $\mathbf{C_{r}}$ denotes the reduced capacitance matrix, which can also be used to
describe a circuit system consisting of two direct-coupled grounded transom
qubits, as shown in Fig.~\ref{fig2}(b) (see Appendix~\ref{A} for details). Here, for illustration purpose,
considering a typical case where all the island capacitors take a same capacitance, i.e., $C_{g1}=C_{g2}=C_{g3}=C_{g4}=C_{g}$ and
both the island capacitance $C_{g}$ and the shunt capacitance $C_{q}$ largely exceed the
coupling capacitances, i.e., $\{C_{g},\,C_{q}\}\gg\{C_{d},\,C_{c1},\,C_{c2}\}$, the
matrix $\mathbf{C_{r}}$ can be approximated by
\begin{equation}
\mathbf{C_{r}}\approx\left(
\begin{array}{ccc}\label{eq3}
 C_{d}  & -\frac{C_{d}}{2} &  -\frac{C_{d}(C_{c1}-C_{c2})}{4C_{g}} \\
  -\frac{C_{d}}{2}  & C_{q}+\frac{C_{g}}{2} & -\frac{C_{c1}+C_{c2}}{4}  \\
  -\frac{C_{d}(C_{c1}-C_{c2})}{4C_{g}} & -\frac{C_{c1}+C_{c2}}{4} & C_{q}+\frac{C_{g}}{2} \\
\end{array}
\right).
\end{equation}
Inspecting $\mathbf{C_{r}}$, one can find that both the matrix
elements $[\mathbf{C_{r}}]_{d,1m}$ and $[\mathbf{C_{r}}]_{d,2m}$, which describe the
coupling between the two qubit modes and the external voltage source, take nonzero values.
This means that although in the original circuit, only one of the two qubits is coupled
to the voltage source, here, both of the two qubits are coupled to the
voltage source simultaneously. As a consequence, after dropping the free modes, the
full circuit shown in Fig.~\ref{fig2}(a) can be transformed into an equivalent circuit shown
in Fig.~\ref{fig2}(b). The most striking result from the equivalence
is that due to the presence of the free modes, a spurious microwave
crosstalk channel can exist, even if the stray coupling between qubits and drive
lines is absent.

To quantify the crosstalk through the spurious channel, we consider
the crosstalk strength defined as $M_{ij}\equiv 20\log_{10}(|\Omega_{i}/\Omega_{j}|)$, expressed in
units of dB. Here, $\Omega_{j}$ denotes the magnitude of the microwave
drive applied to a target qubit (e.g., $Q_{1}$), while $\Omega_{i}$ represents the magnitude of the
crosstalk felt by the other nearby qubit (e.g., $Q_{2}$). For a grounded transmon qubit coupled to an external
voltage source $V_{d}$ via a coupling capacitor $C_{d}$, the magnitude of the
drive can be approximated by $\Omega=C_{d}Q_{\rm zpf}V_{d}/C_{q}$ with $Q_{\rm zpf}=\sqrt{\hbar/2Z_{q}}$ the zero-point
charge fluctuations and $Z_{q}=\sqrt{L/C}$ the qubit impedance \cite{Sank2014}. $L$
and $C$ correspond to the qubit inductance and capacitance, respectively. For illustration
purposes only, we further assume that both qubits have the same qubit inductances and
capacitances, thus in the system shown in Fig.~\ref{fig2}(a), the strength of the spurious
microwave crosstalk from $Q_{1}$ to $Q_{2}$ can be expressed by
\begin{equation}
\begin{aligned}\label{eq4}
&M=20\log_{10}\left(\left|\frac{[\mathbf{C_{r}}]_{d,2m}}{[\mathbf{C_{r}}]_{d,1m}}\right|\right)
=20\log_{10}\left(\mathcal{R}\right),
\end{aligned}
\end{equation}
with the ratio $\mathcal{R}$ given by
\begin{equation}
\begin{aligned}\label{eq4}
\mathcal{R}=\left|\frac{C_{g4}C_{c1}-C_{g3}C_{c2}}{(C_{g3}+C_{g4})(C_{c2}+C_{g2})+C_{g2}(C_{c1}+C_{c2})}\right|.
\end{aligned}
\end{equation}

As shown in Eq.~(\ref{eq4}), the spurious crosstalk only depends
on the qubit circuit parameters, thus acting as an intrinsic
crosstalk channel. Moreover, since in coupled qubit circuits, the
coupling capacitors, e.g., $C_{c1}$ and $C_{c2}$, generally have a rather small capacitance, which is
typical of the order of a few $\rm fFs$ or less, the spurious crosstalk can
be suppressed through increasing the island capacitance. However, depending on
the qubit geometric layout, the island capacitance can take a wide range of
values, typically, ranging from a few fFs to $100\,\rm fF$. Hence, as we will
show below, the spurious crosstalk can become non-negligible, thus limiting
the performance of qubit addressing.

While here we focus on the direct-coupled qubit system, in
Appendices~\ref{B} and~\ref{C}, we also extend the above analysis to
the indirect-coupled qubit systems, including
floating qubits coupled through a grounded bus or a floating
bus. We find that the spurious crosstalk channel disappears for floating qubits coupled via
the grounded bus, while it still exists for qubits coupled via the floating bus.
This opposite conclusions further demonstrate that the presence of the spurious crosstalk
is mediated by the free modes in the qubit circuit. In addition, for
floating qubits coupled by the grounded bus, the analysis also shows that the
grounded bus can also feel the drive applied to the floating qubit, i.e., a spurious
crosstalk channel between the floating qubit and the grounded bus is existent. Thus, we
conclude that for the crosstalk from $Q_{1}$ to $Q_{2}$ in the system shown in Fig.~\ref{fig2}(a), this
spurious crosstalk is in fact mediated by the free mode supported by the
driven qubit $Q_{1}$, and it will still exist even if $Q_{2}$ is a grounded one.

\begin{figure}[tbp]
\begin{center}
\includegraphics[keepaspectratio=true,width=\columnwidth]{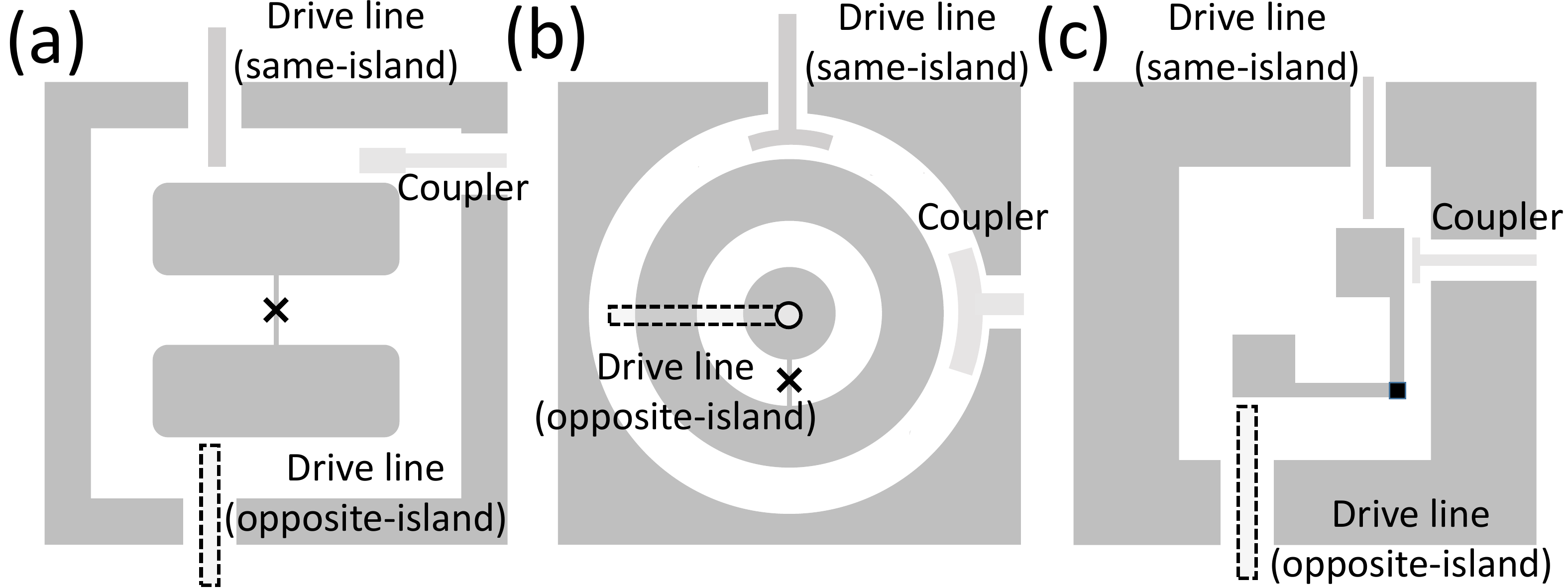}
\end{center}
\caption{Different qubit geometric layouts of floating transmon qubits. The drive line (for single-qubit addressing)
and coupler (for coupling two qubits) can be coupled capacitively to the same superconducting island or two opposite
islands, thus giving rise to two different coupling geometric layouts, i.e., the same-island
layout and the opposite-island layout. (a) The typical floating transmon qubit with symmetric
islands. (b) The coaxial transmon (coaxmons) with asymmetric islands. (c) The floating
merged-element transmon qubit (MET). In (a) and (b), the capacitance between the island and the
ground (island capacitance), in general, largely exceeds the coupling capacitor between qubits (coupling capacitance), while in (c) the island
capacitance can be comparable with or even be smaller than the coupling capacitance.}
\label{fig3}
\end{figure}

\begin{table}
  \caption{Summary of crosstalk ($\mathcal{R}$) in floating transmon qubits with different
  qubit geometric layouts. Here $r\equiv C_{c}/C_{g}$ denotes the ratio of the coupling
  capacitance to the island capacitance, and hereafter we refer to it as the capacitance ratio.}
\renewcommand\arraystretch{2}
  \begin{tabular}{ccc}
  \hline\hline	
    Geometric layout& \begin{tabular}{c} Asymmetric\\($\lambda\neq 1$) \end{tabular} &\begin{tabular}{c} Symmetric\\$(\lambda=1$)\end{tabular} \\

  \hline
    \begin{tabular}{c} Same-island \\ ($C_{c1}=C_{c},\,C_{c2}=0$) \end{tabular} & \begin{tabular}{c} $\displaystyle\frac{C_{c1}}{(\lambda+1)C_{g}+C_{c1}}$ \\ $=\displaystyle\frac{r}{\lambda+1+r}$ \end{tabular} & \begin{tabular}{c} $\displaystyle\frac{C_{c1}}{2C_{g}+C_{c1}}$\\$=\displaystyle\frac{r}{2+r}$ \end{tabular} \\
  \hline	
     \begin{tabular}{c} Opposite-island \\ ($C_{c1}= 0,\,C_{c2}=C_{c}$) \end{tabular} & \begin{tabular}{c} $\displaystyle\frac{C_{c2}}{(\lambda+1)\lambda C_{g}+(2\lambda+1)C_{c2}}$\\$=\displaystyle\frac{r}{\lambda (\lambda+1)+(2\lambda+1)r}$ \end{tabular} & \begin{tabular}{c} $\displaystyle\frac{C_{c2}}{2C_{g}+3C_{c2}}$\\$=\displaystyle\frac{r}{2+3r}$ \end{tabular} \\

  \hline\hline
  \end{tabular}
\label{tabI}
\end{table}

\section{spurious microwave crosstalk in typical floating transmon circuits}\label{SecIII}

The above discussion indicates that the spurious crosstalk channel is intrinsic,
and its strength only depends on the qubit circuit
parameters, i.e., the concrete qubit geometric layout. Here, to study
this geometric dependence, we consider that for the coupled qubit system shown in Fig.~\ref{fig2}(a), $C_{g1}=C_{g3}=C_{g}$
and $C_{g2}=C_{g4}=\lambda C_{g}$, thus giving rise to
\begin{equation}
\begin{aligned}\label{eq5}
\mathcal{R}=\left|\frac{\lambda C_{c1}-C_{c2}}{(\lambda+1)(C_{c2}+\lambda C_{g})+\lambda(C_{c1}+C_{c2})}\right|.
\end{aligned}
\end{equation}
Thus, according to the ratio $\lambda$ of the two
island capacitances (hereafter we refer to it as the island asymmetric ratio), there
are two main island geometric layouts for floating transmon qubits, i.e., symmetry
layout and asymmetry layout. For example, the traditional floating transmon qubits have two
same islands \cite{Paik2020}, as shown in Fig.~\ref{fig3}(a), thus acting as a symmetric one, while for the
coaxial transmon \cite{Rahamim2017}, as shown in Fig.~\ref{fig3}(b), its two islands have different
geometric designs, thus it belongs to the asymmetric one.

Additionally, note that in principle, the drive line and the
coupler (e.g., the capacitors $C_{c1}$ and $C_{c2}$) could be
coupled to one of the two islands or both of the two islands simultaneously, as shown in Fig.~\ref{fig2}(a).
However, for practically implemented floating qubit circuits, as shown
in Fig.~\ref{fig3}, generally, the drive line and the coupler are dominantly coupled
to one of the two islands. Hence, as shown in Fig.~\ref{fig3}, here we consider two coupling
geometric layouts, i.e., the drive line could be coupled
capacitively to the same island or the opposite island with respect to
the coupling island, which is coupled capacitively to the other
nearby qubits. For example, in Fig.~\ref{fig2}(a), when $C_{c1}\neq 0$ and $C_{c2}=0$, the
coupling geometric layout is a same-island one, while for $C_{c1}= 0$ and $C_{c2}\neq0$, it is
an opposite-island case.

\begin{figure}[tbp]
\begin{center}
\includegraphics[keepaspectratio=true,width=\columnwidth]{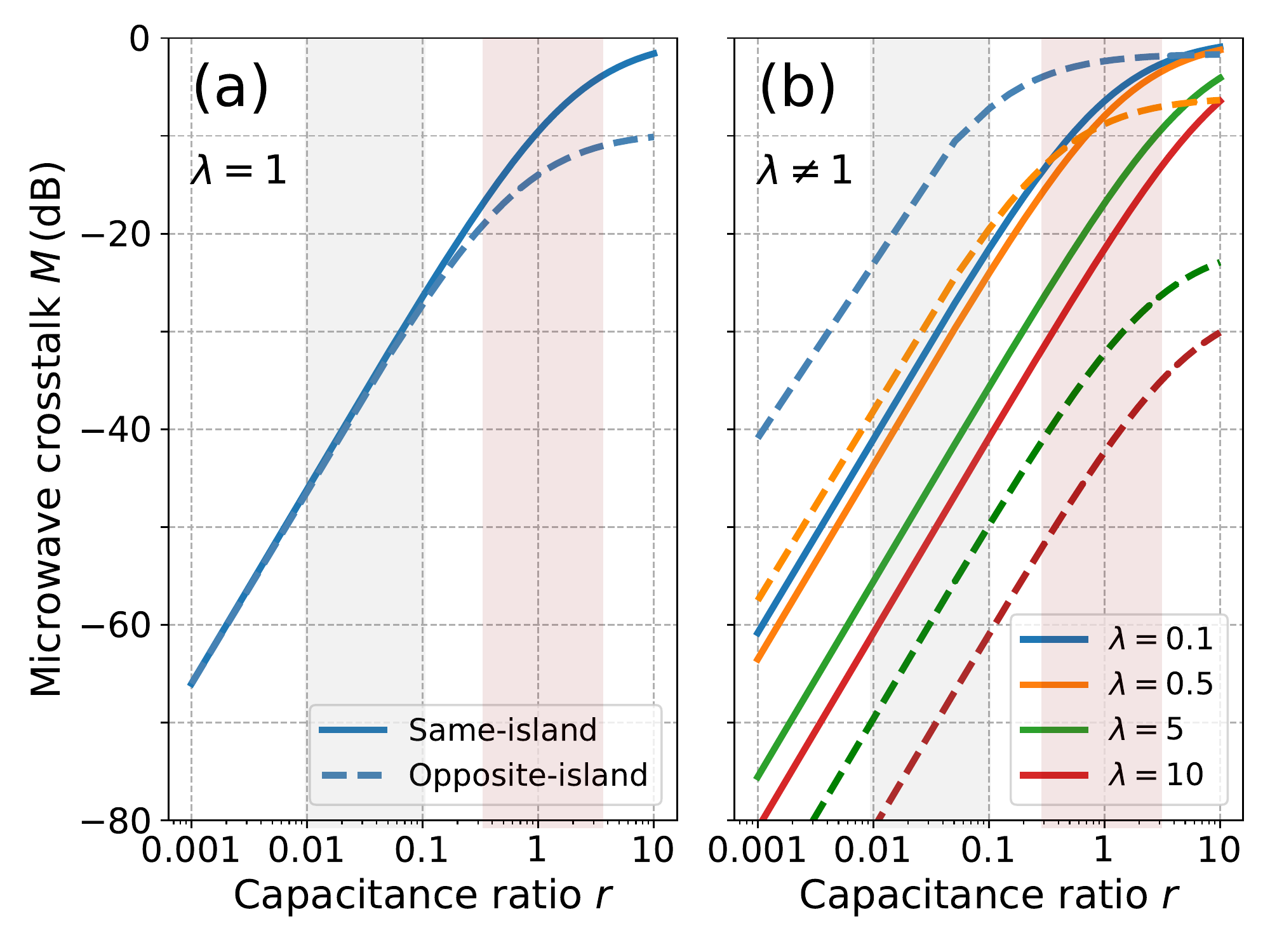}
\end{center}
\caption{The spurious crosstalk for coupled floating transmon qubits with different
circuit geometric layouts. (a) the spurious crosstalk versus the capacitance ratio $r$ for
symmetric qubit geometric layout, i.e., $\lambda=1$. (b) the spurious
crosstalk versus the capacitance ratio $r$ for asymmetric qubit geometric
layout, i.e., $\lambda\neq1$. The solid and dashed lines denote the results for qubit circuits
with the two different coupling geometric layouts, i.e., the same-island layout and opposite-island
layout, respectively. The gray (light gray) area highlights the ratio
band of the typical floating transmon qubits, as shown in Fig.~\ref{fig3}(a) and (b), while the pink (gray) area
for the floating MET, as shown in Fig.~\ref{fig3}(c).}
\label{fig4}
\end{figure}

According to the above mentioned island geometric layout (symmetric v.s. asymmetric) and
coupling geometric layout (same-island v.s. opposite-island), Table~\ref{tabI} lists
the expressions of the crosstalk strengths for four different qubit geometric
layouts. Accordingly, Figure~\ref{fig4} shows the crosstalk strengths as a
function of the capacitance ratio $r$, i.e., the ratio of the
coupling capacitance to the island capacitance. As shown in Fig.~\ref{fig4}(a), for
the traditional floating transmon qubits, where the island capacitance is generally far
larger than the coupling capacitance, thus the spurious crosstalk can be
entirely suppressed below, e.g., $-30\,\rm dB$, regardless of the coupling geometric layout.
However, for the floating merged-element transmon qubit (MET) with a symmetric layout \cite{Mamin2022}, where
the island capacitance can be comparable to or even smaller than the coupling
capacitance, the spurious channel can become the dominated microwave crosstalk source. In this situation,
the magnitude of the spurious crosstalk can even be comparable to that of the
target dive. In addition, compared to the case of the same-island layout, qubits with
the opposite-island layout, especially for the floating MET qubits, can show a less
pronounced spurious crosstalk, typically below $-10\,\rm dB$.

Figure~\ref{fig4}(b) shows the spurious crosstalk strength for qubits with
the asymmetric island layout. One can find that generally larger
island asymmetric ratio $\lambda$ can mitigate the spurious crosstalk, and can
become even more noticeable when taking the opposite-island layout. Thus, for
the coaxial transmon shown in Fig.~\ref{fig3}(b), when the
drive line is coupled to the center island and inter-qubit coupling is realized via
the outer island, the spurious crosstalk could be readily suppressed below $-50\,\rm dB$.
Moreover, while the spurious channel acts as a significant source of the
crosstalk for the MET qubits with the symmetric island
layout, by taking the opposite-island layout and increasing the island asymmetric
ratio, the spurious crosstalk is promising to be pushed below $-25\,\rm dB$.

Given the state-of-the-art values of the microwave crosstalk in multiqubit
quantum processors, typically
ranging from $-25\,\rm dB$ to $-30\,\rm dB$ for neighboring control
lines \cite{Gong2021,Ren2022}, we conclude that:
(i) for the traditional floating transmon qubits with symmetric
island layout, as shown in Fig.~\ref{fig3}(a), the spurious
crosstalk can be ignored safely if the island capacitance largely exceeds
the coupling capacitance. (ii) for the
floating transmon qubits with asymmetric island layout, such as
the coaxial transmon shown in Fig.~\ref{fig3}(b), taking a larger asymmetric ratio
could help to suppress the spurious crosstalk. This suppression
can become more prominent when employing the opposite-island layout.
(iii) for the floating MET qubits with symmetric island layout, as shown
in Fig.~\ref{fig3}(c), the spurious crosstalk is less readily suppressed.
Nevertheless, similar to the coaxial transmon, the combination
of the asymmetric island layout and the opposite-island
layout will hopefully lead to adequate suppression of the
spurious crosstalk.

\section{discussion}\label{SecIV}

Recently, two independent works show that the free mode can mediate an indirect inter-qubit coupling \cite{Sete2021,Yanay2022}. In
Ref.\cite{Sete2021}, Sete \emph{et al.} have demonstrated that for qubits coupled via a floating
tunable coupler (i.e., an additional frequency-tunable floating transmon qubit), the free mode due
to the floating structure of the coupler can induce an indirect inter-qubit coupling, for which
its strength only depends on the circuit capacitance, rather than the coupler frequency. Thus,
combined with the coupler mediated dispersive coupling, tunable coupling between qubits can be realized
by adjusting the coupler frequency \cite{Sete2021,Yan2018}. Similarly, in Ref.\cite{Yanay2022}, Yanay \emph{et al.} have
shown that in an array of floating transmon qubits, the free mode can mediate interactions between
next-nearest neighbor qubits and even beyond next-nearest neighbors. The strength of the
mediated interaction can be engineered via circuit design. Similarly, our
results show that the free mode can also mediate cross-driving, enabling the presence of
an additional intrinsic crosstalk channel. Hence, our analysis can be viewed as complements to
the two earlier works, extending the free-mode mediated interactions
from "quantum regime" (for inter-qubit coupling) to "semi-classical regime" (for spurious
microwave crosstalk). Moreover, by extending to this semi-classical regime, it is
also reasonable to expect that the floating transmon qubit can also relax through the
control lines of its coupled neighbors.

\section{conclusion}\label{SecV}

In summary, by performing circuit analysis of two coupled floating transmon
qubits, we have demonstrated that the free mode, which is supported
by the floating structure of qubits, can induce a spurious microwave
crosstalk channel. Depending on the actual qubit geometric layout, the
spurious crosstalk can become non-negligible, and can even limit the
performance of qubit addressing. Thus, to ensure higher-fidelity qubit
addressing, the spurious crosstalk needs to be carefully considered. To address it, we
have also shown that for various typical floating transmon circuits, the spurious
crosstalk can be largely suppressed through circuit design.

Although our present analysis of the spurious crosstalk focuses on the
floating transmon qubits, we expect that the analysis and many of the results
may also be applied to other types of superconducting circuits with floating islands, such as
flux qubits \cite{Orlando1999} and fluxonium qubits \cite{Manucharyan2009}.

\begin{acknowledgments}

We acknowledge helpful discussions with Huihai Zhao and Zhenyu Mi. This work
was supported by the National Natural Science Foundation of
China (Grants  No.11890704, No.12104055, No.12004042, No.12104056), the Beijing
Natural Science Foundation (Grant No.Z190012), and the
Key-Area Research and Development Program of Guang Dong Province (Grant No. 2018B030326001).

\end{acknowledgments}

\appendix

\section{Direct Capacitor}\label{A}
Here, we consider two floating qubits coupled via direct capacitors, as
shown in Fig.~\ref{fig2}(a). The Lagrangian of the circuit can be
expressed in terms of the node flux variables $\Phi_{n}$ with $n=\{d,\,1,\,2,\,3,\,4\}$, as
denoted in Fig.~\ref{fig2}(a), and is given as \cite{Vool2017}
\begin{equation}
\begin{aligned}\label{eqa1}
\mathcal{L'}=\frac{1}{2}\mathbf{\dot{\Phi'}}^{T} \mathbf{C'}\mathbf{\dot{\Phi'}}+E_{J1}\cos(\phi_{1}-\phi_{2})+E_{J2}\cos(\phi_{3}-\phi_{4})
\end{aligned}
\end{equation}
where where $E_{J1}$ and $E_{J2}$ are the Josephson energies, $\phi_{n}=2\pi\Phi_{n}/\Phi_{0}$ with $\Phi_{0}$ the magnetic flux quantum, $\mathbf{\Phi'}\equiv(\Phi_{d}\,\Phi_{1}\,\Phi_{2}\,\Phi_{3}\,\Phi_{4})^{T}$, and $\mathbf{C'}$
the capacitance matrix, given as
\begin{equation}
\mathbf{C'}=\left(
\begin{array}{ccccc}\label{eqa2}
C_{d} &-C_{d} & 0& 0 &0\\
-C_{d} & C_{\Sigma_{1}} & -C_{q}& -C_{c1} &0\\
0& -C_{q} & C_{\Sigma_{2}} & 0 &-C_{c2}\\
0& -C_{c1} & 0& C_{\Sigma_{3}} &-C_{q}\\
0& 0 & -C_{c2}& -C_{q} & C_{\Sigma_{4}}\\
\end{array}
\right)
\end{equation}
with
\begin{equation}
\begin{aligned}\label{eqa3}
& C_{\Sigma_{1}}=C_{q}+C_{g1}+C_{c1}+C_{d},
\\&C_{\Sigma_{2}}=C_{q}+C_{g2}+C_{c2},
\\&C_{\Sigma_{3}}=C_{q}+C_{g3}+C_{c1},
\\&C_{\Sigma_{4}}=C_{q}+C_{g4}+C_{c2}.
\end{aligned}
\end{equation}

To identify and remove the free modes in the circuit, we consider the following
transformation of the node flux variables with respect to the transformation matrix \cite{Yanay2020}
\begin{equation}
\textbf{S}=\left(
\begin{array}{ccccc}\label{eqa4}
1 &0 & 0& 0 &0\\
0 & 1 & 1& 0 &0\\
0& 1 & -1& 0 &0\\
0& 0 & 0& 1 &1\\
0& 0 & 0& 1 &-1\\
\end{array}
\right).
\end{equation}
After performing the transformation, the circuit Lagrangian now reads
\begin{equation}
\begin{aligned}\label{eqa5}
\mathcal{L}=\frac{1}{2}\mathbf{\dot{\Phi}}^{T} \mathbf{C}\mathbf{\dot{\Phi}}+E_{J1}\cos(\phi_{1m})+E_{J2}\cos(\phi_{2m}).
\end{aligned}
\end{equation}
Here, the transformed node flux variables is $\mathbf{\Phi}=S\mathbf{\Phi'}=(\Phi_{d}\,\Phi_{1p}\,\Phi_{1m}\,\Phi_{2p}\,\Phi_{2m})^{T}$,
where $\Phi_{1p(m)}=\Phi_{1}\pm\Phi_{2}$ and $\Phi_{2p(m)}=\Phi_{3}\pm\Phi_{4}$,
and the corresponding capacitance matrix is now given as $\mathbf{C}=S^{-1}\mathbf{C'}S^{-1}$. Then, the
circuit Hamiltonian can be expressed as \cite{Vool2017}
\begin{equation}
\begin{aligned}\label{eqa6}
H=\sum_{j}Q_{j}\dot{\Phi}_{j}-E_{J1}\cos(\phi_{1m})-E_{J2}\cos(\phi_{2m}),
\end{aligned}
\end{equation}
where $Q_{j}=\partial\mathcal{L}/\partial \dot{\Phi}_{j}$ denotes the charge variable, which
is conjugated to the node flux variable $\Phi_{j}$ with $j=\{d,\,1p,\,1m,\,2p,\,2m\}$. Here,
the modes associated with $Q_{1p}$ and $Q_{2p}$, and which don't have any potential energies in
the circuit Hamiltonian, are the free modes \cite{Kerman2020,Ding2021,Long2020}.

As discussed in previous works, the free modes actually don't participate in the circuit
dynamics \cite{Kerman2020,Ding2021,Long2020}. Hence, one can drop the terms associated with the free modes in the circuit
Hamiltonian in Eq.~(\ref{eqa6}), giving rise to
\begin{equation}
\begin{aligned}\label{eqa7}
H_{r}=\frac{1}{2}\mathbf{Q_{r}}^{T} \mathbf{C_{r}}^{-1}\mathbf{Q_{r}}-E_{J1}\cos(\phi_{1m})-E_{J2}\cos(\phi_{2m})
\end{aligned}
\end{equation}
where $\mathbf{Q_{r}}=(Q_{d}\,Q_{1m}\,Q_{2m})^{T}$, $\mathbf{C_{r}}$ denotes the reduced capacitance
matrix, which can also be used to describe a system consisting of two direct-coupled grounded
transom qubits, as shown in Fig.~\ref{fig2}(b). Here, for illustration purpose and to avoid the
extremely cumbersome expression of $\mathbf{C_{r}}$, we consider that all the island capacitor take the
same capacitance, i.e., $C_{g1}=C_{g2}=C_{g3}=C_{g4}=C_{g}$. In this case, the expression
of $\mathbf{C_{r}}$ is given by (since $\mathbf{C_{r}}$ is a symmetric matric, hereafter, the elements in
the lower triangular parts of the capacitance matrix, denoted by $\blacksquare$, are not
given explicitly)
\begin{widetext}
\begin{equation}
\mathbf{C_{r}}=\left(
\begin{array}{ccc}\label{eqa8}
 \tilde{C}_{\Sigma d}   & -\frac{C_{d}C_{g}(C_{c1}+2C_{g}+3C_{c2})}{K}& -\frac{C_{d}C_{g}(C_{c1}-C_{c2})}{K} \\
 \blacksquare & \tilde{C}_{\Sigma q1} & -\frac{C_{g}(C_{d}+C_{g})C_{c2}+C_{c1} \left(C_{g}^{2}+4C_{c2}C_{g}+C_{d}C_{c2}\right)}{K}  \\
  \blacksquare & \blacksquare & \tilde{C}_{\Sigma q2} \\
\end{array}
\right),
\end{equation}
\end{widetext}
where
\begin{equation}
\begin{aligned}\label{eqa9}
K=C_{c1}(C_{d}+4C_{g})+4C_{g}(C_{g}+C_{c2})+C_{d}(2C_{g}+C_{c2}),
\end{aligned}
\end{equation}

\begin{equation}
\begin{aligned}\label{eqa10}
\tilde{C}_{\Sigma d}=\frac{4C_{d}C_{g} (C_{c1}+C_{g}+C_{c2})}{K}
\end{aligned}
\end{equation}

\begin{equation}
\begin{aligned}\label{eqa11}
\tilde{C}_{\Sigma q1}=&\big[C_{d}\left(2C_{g}^{2}+(2C_{q}+3C_{c2})C_{g}+C_{q}C_{c2}\right)
\\&+C_{g} \left(2C_{g}^{2}+(4C_{q}+3C_{c2})C_{g}+4 C_{q}C_{c2}\right)
\\&+C_{c1}C_{d} (C_{g}+C_{q}+C_{c2})
\\&+C_{c1}C_{g} (3C_{g}+4(C_{q}+C_{c2}))\big]/K
\end{aligned}
\end{equation}

\begin{equation}
\begin{aligned}\label{eqa12}
\tilde{C}_{\Sigma q2}= &\big[C_{d} \left(C_{g}^{2}+(2C_{q}+C_{c2})C_{g}+C_{q}C_{c2}\right)
\\&+C_{g}\left(2C_{g}^{2}+(4C_{q}+3C_{c2}) C_{g}+4 C_{q}C_{c2}\right)
\\&+C_{c1}C_{d} (C_{g}+C_{q}+C_{c2})
\\&+C_{c1}C_{g} (3C_{g}+4(C_{q}+C_{c2}))\big]/K
\end{aligned}
\end{equation}

When the island capacitance $C_{g}$ and the shunt capacitance $C_{q}$ is far larger than the coupling capacitance, i.e., $\{C_{g},\,C_{q}\}\gg\{C_{d},\,C_{c1},\,C_{c2}\}$, the reduced capacitance matrix $C_{r}$ can be approximated by
\begin{equation}
\mathbf{C_{r}}\approx\left(
\begin{array}{ccc}\label{eqa13}
 C_{d}  & -\frac{C_{d}}{2} &  -\frac{C_{d}(C_{c1}-C_{c2})}{4C_{g}} \\
  -\frac{C_{d}}{2}  & C_{q}+\frac{C_{g}}{2} & -\frac{C_{c1}+C_{c2}}{4}  \\
  -\frac{C_{d}(C_{c1}-C_{c2})}{4C_{g}} & -\frac{C_{c1}+C_{c2}}{4} & C_{q}+\frac{C_{g}}{2} \\
\end{array}
\right)
\end{equation}

\begin{figure}[tbp]
\begin{center}
\includegraphics[keepaspectratio=true,width=\columnwidth]{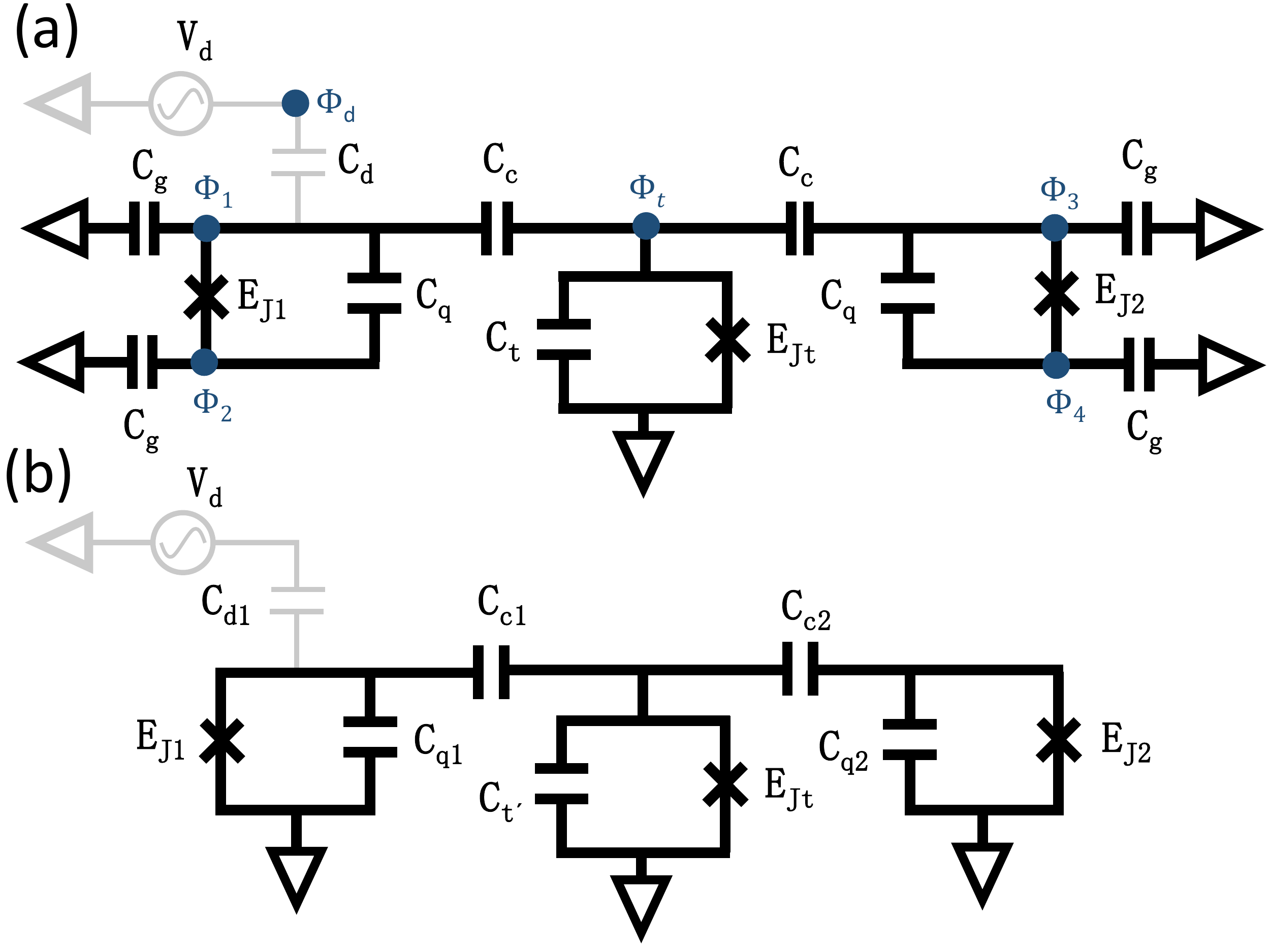}
\end{center}
\caption{(a) Schematic circuit diagram for two floating transmon qubits ($Q_{1}$ and $Q_{2}$) coupled via a
grounded bus ($Q_{B}$), where a dedicated
drive line (left) is coupled capacitively to one of the two qubits, i.e., $Q_{1}$. After removing free
modes, the circuit can be transformed into an equivalent circuit shown in (b), where two
grounded transmon qubits are coupled via a grounded bus. Note here that unlike the results
shown in Fig.~\ref{fig2}, the spurious crosstalk between the two qubits disappears, giving rise
to $C_{d2}=0$. This can be explained by the fact that (i) there is no direct coupling between the two qubits; (2)
the grounded bus does not support the presence of free mode; Thus, there are no free modes to mediate the
spurious crosstalk between the two qubits.}
\label{fig5}
\end{figure}

\section{Grounded Bus}\label{B}

Here, as shown in Fig.~\ref{fig5}(a), we consider that two floating transmon qubits coupled via a
grounded bus. Following the same procedure given in Appendix~\ref{A}, the circuit in Fig.~\ref{fig5}(a)
can be reduced to the circuit in Fig.~\ref{fig5}(b), where two grounded transmon qubits are coupled via
a grounded bus. After removing free modes, the reduced capacitance matrix of the current
circuit is given by (here, the charge variables are $\mathbf{Q_{r}}=(Q_{d}\,Q_{1m}\,Q_{t}\,Q_{2m})^{T}$,
and the corresponding flux variables are $\mathbf{\Phi_{r}}=(\Phi_{d}\,\Phi_{1m}\,\Phi_{t}\,\Phi_{2m})^{T}$
with $\Phi_{1m}=\Phi_{1}-\Phi_{2}$ and $\Phi_{2m}=\Phi_{3}-\Phi_{4}$)

\begin{equation}
\mathbf{C_{r}}=\left(
\begin{array}{cccc}\label{eqb1}
 \tilde{C}_{\Sigma d} & -\frac{C_{d}C_{g}}{C_{d}+C_{c}+2C_{g}} & -\frac{C_{d}C_{c}}{C_{d}+C_{c}+2 C_{g}} & 0 \\
 \blacksquare & \tilde{C}_{\Sigma q1} & -\frac{C_{c}C_{g}}{C_{d}+C_{c}+2C_{g}} & 0 \\
 \blacksquare & \blacksquare & \tilde{C}_{\Sigma t} & -\frac{C_{c} C_{g}}{C_{c}+2 C_{g}}\\
 \blacksquare & \blacksquare & \blacksquare & \tilde{C}_{\Sigma q2} \\
\end{array}
\right),
\end{equation}
where
\begin{equation}
\begin{aligned}\label{eqb2}
&\tilde{C}_{\Sigma d}=\frac{C_{d}(C_{c}+2C_{g})}{C_{d}+C_{c}+2C_{g}},
\\&\tilde{C}_{\Sigma q1}=\frac{C_{d} (C_{g}+C_{q})+C_{c} (C_{g}+C_{q})+C_{g} (C_{g}+2 C_{q})}{C_{d}+C_{c}+2C_{g}},
\\&\tilde{C}_{\Sigma q2}=\frac{C_{c} (C_{g}+C_{q})+C_{g} (C_{g}+2C_{q})}{C_{c}+2C_{g}},
\end{aligned}
\end{equation}
and
\begin{equation}
\begin{aligned}\label{eqb3}
\tilde{C}_{\Sigma t}=&\frac{(C_{c}+2C_{g}) (2C_{g}C_{t}+C_{f}(4 C_{g}+C_{t}))}{(C_{c}+2C_{g}) (C_{d}+C_{c}+2C_{g})}
\\&+\frac{C_{d} \left(C_{c}^{2}+(4C_{g}+C_{t})C_{c}+2C_{g}C_{t}\right)}{(C_{c}+2C_{g}) (C_{d}+C_{c}+2C_{g})}.
\end{aligned}
\end{equation}

When considering that $\{C_{g},\,C_{q},\,C_{t}\}\gg\{C_{d},\,C_{c}\}$, the above
matrix can be approximated by
\begin{equation}
\mathbf{C_{r}}\approx\left(
\begin{array}{cccc}\label{eqb4}
 C_{d}  & -\frac{C_{d}}{2} &  -\frac{C_{d}C_{c}}{2C_{g}} & 0\\
  \blacksquare  & C_{q}+\frac{C_{g}}{2} & -\frac{C_{c}}{2} & 0 \\
 \blacksquare & \blacksquare & C_{t} & -\frac{C_{c}}{2} \\
  \blacksquare & \blacksquare & \blacksquare & C_{q}+\frac{C_{g}}{2} \\
\end{array}
\right)
\end{equation}

As shown in Eq.~(\ref{eqb1}), for qubits coupled via the grounded bus, the spurious crosstalk
disappears, i.e., the matrix element, which represents the coupling between the
drive source $V_{d}$ and the $Q_{2}$,  takes the value of $0$, i.e., $[C_{r}]_{d,2m}=-C_{d2}=0$. This
is to be expected since the grounded bus does not support the presence of free mode, thus there
are no free modes to mediate the spurious crosstalk between the two qubits.

In addition, note that the matrix element $[C_{r}]_{d,t}$ gets a nonzero value, as shown
in Eq.~(\ref{eqb1}). This means that there exists a spurious crosstalk channel
between the qubit $Q_{1}$ and the grounded bus $Q_{t}$ (in Fig.~\ref{fig5}(b), the corresponding virtual
drive line is not presented explicitly). Thus, we can conclude that this
spurious crosstalk channel is mediated by the free mode supported by the
qubit $Q_{1}$.

\section{Floating Bus}\label{C}

\begin{figure}[tbp]
\begin{center}
\includegraphics[keepaspectratio=true,width=\columnwidth]{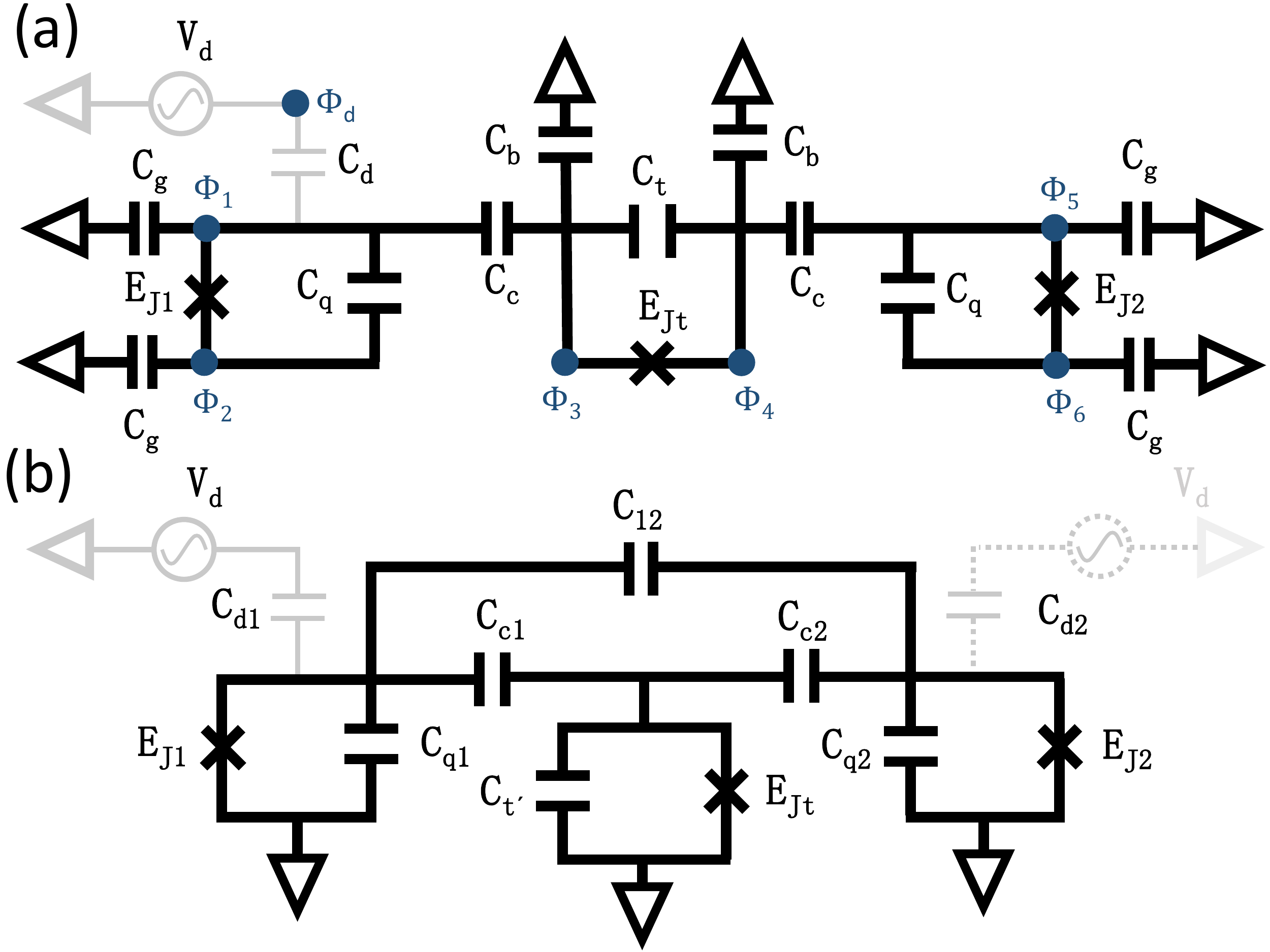}
\end{center}
\caption{(a) Schematic circuit diagram for two floating transmon qubits ($Q_{1}$ and $Q_{2}$) coupled via a
floating bus ($Q_{B}$), where a dedicated
drive line (left) is coupled capacitively to one of the two qubits, i.e., $Q_{1}$. After removing the free
modes supported by the floating structure of qubits, the circuit can be transformed into an equivalent circuit shown in (b), where two
grounded transmon qubits are coupled via a grounded bus. Here, the spurious crosstalk between the
two qubits exists due to the presence of the free mode supported by the floating bus.}
\label{fig6}
\end{figure}

Here, as shown in Fig.~\ref{fig6}(a), we consider that two floating transmon qubits coupled via a
floating bus \cite{Sete2021}. Following the same procedure given in Appendix~\ref{A}, the circuit in Fig.~\ref{fig6}(a)
can be transformed into an equivalent circuit shown in Fig.~\ref{fig6}(b), where two grounded transmon qubits are coupled via
a coupler circuit combining a grounded bus and a capacitor \cite{Sete2021,Yan2018}. Accordingly, the reduced
capacitance matrix of the present circuit is given by (here, the charge variables
are $\mathbf{Q_{r}}=(Q_{d}\,Q_{1m}\,Q_{t}\,Q_{2m})^{T}$, and the corresponding flux variables are $\mathbf{\Phi_{r}}=(\Phi_{d}\,\Phi_{1m}\,\Phi_{t}\,\Phi_{2m})^{T}$
with $\Phi_{1m}=\Phi_{1}-\Phi_{2}$, $\Phi_{t}=\Phi_{3}-\Phi_{4}$, and $\Phi_{2m}=\Phi_{5}-\Phi_{6}$)

\begin{widetext}
\begin{equation}
\mathbf{C_{r}}=\left(
\begin{array}{cccc}\label{eqc1}
  \tilde{C}_{\Sigma d} & -\frac{C_{d}C_{g}(2C_{b} (C_{c}+2C_{g})+C_{c} (C_{c}+4C_{g}))}{K} & -\frac{C_{c}C_{d} (2C_{c} C_{g}+C_{b} (C_{c}+2 C_{g}))}{K} & -\frac{C_{c}^{2} C_{d} C_{g}}{K} \\
  \blacksquare &\tilde{C}_{\Sigma q1} & -\frac{C_{c}C_{g} (2C_{c} C_{g}+C_{b} (C_{c}+2 C_{g}))}{K} & -\frac{C_{c}^{2}C_{g}^{2}}{K} \\
 \blacksquare & \blacksquare & \tilde{C}_{\Sigma t} & \frac{C_{c}C_{g} (C_{c} (C_{d}+2C_{g})+C_{b} (C_{c}+C_{d}+2C_{g}))}{K} \\
 \blacksquare & \blacksquare & \blacksquare & \tilde{C}_{\Sigma q2} \\
\end{array}
\right),
\end{equation}
\end{widetext}
where
\begin{equation}
\begin{aligned}\label{eqc2}
K=&2C_{b}(C_{c}+2C_{g}) (C_{c}+C_{d}+2C_{g})
\\&+C_{c}(4C_{g} (C_{d}+2C_{g})+C_{c}(C_{d}+4C_{g})),
\end{aligned}
\end{equation}

\begin{equation}
\begin{aligned}\label{eqc3}
&\tilde{C}_{\Sigma d}=\frac{2C_{d} (C_{c}+2C_{g}) (2C_{c}C_{g}+C_{b} (C_{c}+2C_{g}))}{K},
\end{aligned}
\end{equation}

\begin{equation}
\begin{aligned}\label{eqc4}
\tilde{C}_{\Sigma q1}=&\big[2C_{b}(C_{c}+2C_{g})(C_{c} (C_{g}+C_{q})+C_{d} (C_{g}+C_{q})
\\&+C_{g}(C_{g}+2C_{q}))+C_{c}(4C_{g}(C_{d} (C_{g}+C_{q})
\\&+C_{g} (C_{g}+2C_{q}))+C_{c} (C_{d} (C_{g}+C_{q})
\\&+C_{g} (3C_{g}+4C_{q})))\big]/K,
\end{aligned}
\end{equation}

\begin{equation}
\begin{aligned}\label{eqc5}
\tilde{C}_{\Sigma t}=&\big[(C_{c}+2C_{g}) (C_{c}+C_{d}+2C_{g})C_{b}^{2}
\\&+((C_{d}+4C_{g}+2C_{t})C_{c}^{2}+2(4C_{g} (C_{g}+C_{t})
\\&+C_{d} (2 C_{g}+C_{t})) C_{c}+4C_{g} (C_{d}+2C_{g})C_{t})C_{b}
\\&+C_{c} (4C_{g} (C_{d}+2C_{g})C_{t}+C_{c} (4C_{g} (C_{g}+C_{t})
\\&+C_{d} (2 C_{g}+C_{t})))\big]/K,
\end{aligned}
\end{equation}

\begin{equation}
\begin{aligned}\label{eqc6}
\tilde{C}_{\Sigma q2}=&\big[2C_{b} (C_{c}+C_{d}+2C_{g}) (C_{c} (C_{g}+C_{q})
\\&+C_{g} (C_{g}+2C_{q}))+C_{c} (2C_{g} (C_{d}+2C_{g}) (C_{g}+2 C_{q})
\\&+C_{c} (C_{d} (C_{g}+C_{q})+C_{g} (3C_{g}+4C_{q})))\big]/K.
\end{aligned}
\end{equation}

When considering that $\{C_{g},\,C_{q},\,C_{t},\,C_{b}\}\gg\{C_{d},\,C_{c}\}$, the above
matrix can be approximated by
\begin{equation}
\mathbf{C_{r}}\approx\left(
\begin{array}{cccc}\label{eqc7}
 C_{d}  & -\frac{C_{d}}{2} &  -\frac{C_{d}C_{c}}{4C_{g}} & -\frac{C_{c}^{2}C_{d}}{8C_{b}C_{g}}\\
  \blacksquare  & C_{q}+\frac{C_{g}}{2} & -\frac{C_{c}}{4} & -\frac{C_{c}^{2}}{8C_{b}} \\
 \blacksquare & \blacksquare & C_{t}+\frac{C_{b}}{2} & -\frac{C_{c}}{4} \\
  \blacksquare & \blacksquare & \blacksquare & C_{q}+\frac{C_{g}}{2} \\
\end{array}
\right).
\end{equation}

As shown in Eq.~(\ref{eqc1}), for qubits coupled via the floating bus, the spurious crosstalk between the
two qubits exists due to the presence of the free mode supported by the floating bus. Moreover, as
demonstrated in previous works \cite{Sete2021,Yanay2022}, the free mode, which is supported by the
floating bus, can also mediate an indirect coupling between the two qubits.

\end{document}